\documentclass[aps,preprint,nofootinbib]{revtex4}
\usepackage{graphicx}
\usepackage{epsfig}

\begin{document}

\def\beq{\begin{eqnarray}}
\def\eeq{\end{eqnarray}}
\def\non{\nonumber}
\def\la{\langle}
\def\ra{\rangle}
\def\Un{{\cal U}}
\def\Mbar{\overline{M^0}}
\def\Bmixing{B^0-\overline{B^0}}
\def\Dmixing{D^0-\overline{D^0}}

\def\pr{{Phys. Rev.}~}
\def\prl{{ Phys. Rev. Lett.}~}
\def\pl{{ Phys. Lett.}~}
\def\npb{{ Nucl. Phys. B}~}
\def\epjc{{ Eur. Phys. J. C}~}


\title{ Unparticle Physics Effects on $D^0-\overline {D^0}$ Mixing }

\author{ Xue-Qian Li\footnote{Email: lixq@nankai.edu.cn} and
         Zheng-Tao Wei\footnote{Email: weizt@nankai.edu.cn} }

\affiliation{ Department of Physics, Nankai University, Tianjin
  300071, China }

\begin{abstract}

\noindent The mixing of $K^0-\overline{K^0}$, $D^0-\overline{D^0}$
and $B_{(s)}^0-\overline{B^0_{(s)}}$ provides a sensitive probe to
explore new physics beyond the Standard Model. The scale invariant
unparticle physics recently proposed by Georgi can induce
flavor-changing neutral current and contribute to the mixing at tree
level. We investigate the unparticle effects on $B^0-\overline{B^0}$
and $D^0-\overline{D^0}$ mixing. Especially, the newly observed
$D^0-\overline{D^0}$ mixing sets the most stringent constraints on
the coupling of the unparticle to quarks.

\end{abstract}


\maketitle

\section{Introduction}

It is well-known that scale invariance is broken by
renormalization and dimensional parameters in quantum field
theories. The concept of scale invariance (and more generally the
conformal symmetry) may still play an important role in high
energy physics. For an asymptotically free theory, such as QCD,
the scale invariance is recovered in the high energy limit. In the
concerned practical physics processes of high energies, breaking
of scale invariance can be systematically incorporated in the
anomalous dimensions of operators using the renormalization group
method \cite{Conformal}. It is indicated that the scale invariance
in the infrared region may be quite different and less known
\cite{BZ}. But the idea of scale invariance is so simple and
attractive that there is no a $priori$ to repel it from our world.

In \cite{Georgi1}, Georgi proposed that a scale invariant stuff
contains no particle, but the so-called unparticle. The unparticle
possesses some properties which are different from that of ordinary
particles. The first aspect is that it has a non-trivial scale
dimension $d_\Un$. The dimension of unparticle is in general
fractional rather than an integral number (the dimension for a
fermion is half-integral). The fractional dimension must come from
some complicated dynamics whose details are unknown at present.
Another aspect is that the free unparticle has no definite mass.
That means that the Lorentz-invariant four-momentum square $P^2$ is
not fixed for a real unparticle. Georgi observed that unparticle
with scale dimension looks like a non-integral number $d_{\Un}$ of
invisible massless particles \cite{Georgi1}. To be consistent with
the present experimental observations, the coupling of unparticle to
the ordinary Standard Model (SM) matter must be sufficiently weak.
However, it may be relevant to the TeV physics and might be explored
at the LHC and ILC. The interactions between the unparticle and the
SM particles are described in the framework of low energy effective
theory and lead to various interesting phenomena. There have been
some phenomenological explorations on possible observable effects
caused by unparticles \cite{Georgi1,Georgi2,CKY,LZ,CG,Liao,DY,ACG}.

The mixing of $K^0-\overline{K^0}$, $D^0-\overline{D^0}$ and
$B_{(s)}^0-\overline{B^0_{(s)}}$ is of fundamental importance to
test the SM and explore new physics beyond the SM. In the scenarios
of new physics, there may exist a flavor-changing neutral current
(FCNC) to result in such a mixing which can only be realized via
loops in the framework of the SM. Thus this observable could be
sensitive to new physics effects. In fact, many authors used to
explore evidence of new physics in $\Bmixing$ (or $B^0_s-\overline{
B^0_s}$) mixing because data about the mixing have been available
for a long while. In the proposed scenario \cite{Georgi1}, the
unparticle can couple to different flavors of quarks and induce FCNC
even at tree level as long as the unparticle is neutral. Thus it
will cause new contributions to the particle-antiparticle mixing,
$\Bmixing$, $\Dmixing$ mixing. Generally, based on physics
conjecture, the energy scale concerning unparticle is high that it
should cause smaller influence on the $K^0-\overline{K^0}$ mixing,
especially the SM contribution to the mixing obviously dominates.
The unparticle effects on $B^0_{(s)}-\overline{B^0_{(s)}}$ mixing
had been studied in \cite{LZ,CG} roughly. Since the
$B^0_{(s)}-\overline{B^0_{(s)}}$ mixing parameter $x_{B_{d(s)}}$ is
large and generally the contributions from the SM dominate, and the
new physics effect if it exists, is less important, thus the
observable is not so sensitive to the new physics. Whereas for the D
system, the SM contribution is confirmed to be sufficiently small,
and the $\Dmixing$ mixing parameter (the SM prediction is
$x_{D}<10^{-3}$ \cite{FGLP}) must not be measured by the present
experiments, if there is no new physics. By contraries, if sizable
mixing is measured, new physics should exist and make main
contributions. It is interesting that recently the $\Dmixing$ has
indeed been measured by the Babar and Belle collaborations
\cite{Babar,Belle}, which may be a signature of existence of new
physics. He and Valencia \cite{He} suggested that the mixing is due
to the FCNC in the up-type-quark sector for non-universal $Z'$ model
and obtained constraints on the model parameters by fitting the
data. Instead, we propose that the unparticle scenario is the new
physics which is responsible for the observable $\Dmixing$ mixing.

In this study, we will investigate the effects of the unparticle
physics on the neutral meson mixing including
$B^0_{(s)}-\overline{B^0_{(s)}}$, $\Dmixing$ and
$K^0-\overline{K^0}$ mixing and constrain the coupling parameter of
the concerned interactions between the unparticles and the SM
quarks.

\section{$M^0-\Mbar$ mixing in unparticle physics}

We start with a brief review about the unparticle scenario. It is
assumed that the scale invariant unparticle fields emerge below an
energy scale $\Lambda_\Un$ which is at the order of TeV
\cite{Georgi1}. The interactions of the unparticle with the SM
particle are described by a low energy effective theory. For our
purpose, the coupling of unparticle to quarks is given by following
the standard strategy to construct effective interactions as
 \beq
 \frac{c_S^{q'q}}{\Lambda_\Un^{d_\Un}}\bar q'\gamma_{\mu}(1-\gamma_5)
  q\partial^\mu O_\Un +\frac{c_V^{q'q}}{\Lambda_\Un^{d_\Un-1}}\bar
  q'\gamma_{\mu}(1-\gamma_5)qO_\Un^\mu+h.c. .
 \eeq
where $O_\Un$ and $O_\Un^{\mu}$ denote the scalar and vector
unparticle fields, respectively. The $c_S^{q'q}$ and $c_V^{q'q}$ are
dimensionless coefficients and they depend on different flavors in
general. If the $q$ and $q'$ belong to the same up- or down-type
quark sectors, the above effective interactions may induce FCNC
transitions and provide new physics contribution to the neutral
meson mixing. In order to simplify the phenomenological analysis, we
use the same coefficient for all flavors, $c_S^{q'q}\to c_S$ and
$c_V^{q'q}\to c_V$. Relaxing this restriction does not change our
conclusions.

In this study, we are only interested in the effects of the
unparticle field which serves as an intermediate agent in the FCNC
transition, thus it only appears as a propagator with momentum $P$
and scale dimension $d_\Un$.  The propagator for the scalar
unparticle field is given by \cite{Georgi2,CKY}
 \beq
 \int d^4 x e^{iP\cdot x}\la 0 |TO_\Un(x)O_\Un(0)|0\ra &=&
   i\frac{A_{d_\Un}}{2{\rm sin}(d_\Un\pi)}\frac{1}{(P^2+i\epsilon)^{2-d_\Un}}
   e^{-i(d_\Un-2)\pi},
 \eeq
where
 \beq
 A_{d_\Un}=\frac{16\pi^{5/2}}{(2\pi)^{2d_\Un}}\frac{\Gamma(d_\Un+1/2)}
  {\Gamma(d_\Un-1)\Gamma(2d_\Un)}.
 \eeq
The function ${\rm sin}(d_\Un\pi)$ in the denominator implies that
the scale dimension $d_\Un$ cannot be integral for $d_\Un>1$ in
order to avoid singularity. The phase factor $e^{-i(d_\Un-2)\pi}$
provides a CP conserving phase which produces peculiar interference
effects in high energy scattering processes \cite{Georgi2},
Drell-Yan process \cite{CKY} and CP violation in B decays \cite{CG}.
The propagator for the vector unparticle is similarly given by
 \beq
 \int d^4 x e^{iP\cdot x}\la 0 |TO^{\mu}_\Un(x)O^{\nu}_\Un(0)|0\ra &=&
  i\frac{A_{d_\Un}}{2{\rm sin}(d_\Un\pi)}\frac{-g^{\mu\nu}+P^{\mu}P^{\nu}/P^2}
  {(P^2+i\epsilon)^{2-d_\Un}}e^{-i(d_\Un-2)\pi},
 \eeq
where the transverse condition $\partial_\mu O_\Un^\mu=0$ is used.

\begin{figure}[!htb]
\begin{center}
\begin{tabular}{cc}
\includegraphics[width=14cm]{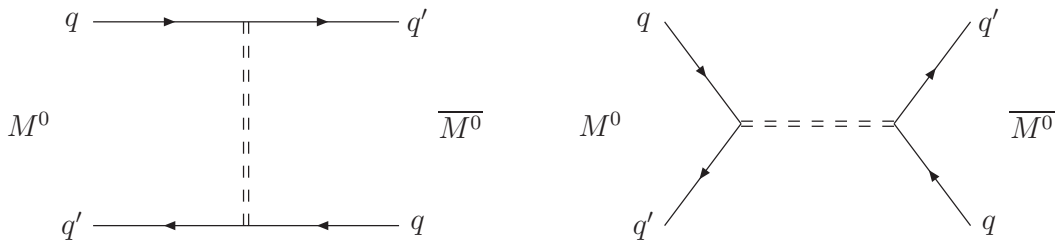}
\end{tabular}
\end{center}
\label{figure} \caption{ The $M^0-\overline{M^0}$ mixing in
unparticle physics. The double dashed lines represent the unparticle
fields.} \label{fig1}
\end{figure}

The neutral meson is denoted by $M^0(q\bar q')$ and its antiparticle
$\Mbar (q' \bar q)$. The mixing occurs via a transition $q\bar q'\to
q' \bar q$ at the quark level. In the SM, these FCNC processes can
only be realized at loop orders. The lowest contribution which
results in the $M^0-\overline{M^0}$ mixing is the box diagrams. With
the unparticle scenario, the FCNC transitions can occur at tree
level and they are depicted in Fig. \ref{fig1}. The double dashed
lines represent the exchanged unparticle fields. There are two
diagrams corresponding to t- and s-channel unparticle-exchanges
which contribute to the $M^0-\overline{M^0}$ mixing.

The $M^0-\Mbar$ mixing is usually described by two parameters: the
mass difference $\Delta m_M$ and width difference $\Delta
\Gamma_M$. The unparticle physics modifies $\Delta m_M$ and thus
changes the SM predictions. For the heavy mesons $B_d,~B_s,~D$,
the mass difference $\Delta m_M$ is related to the mixing matrix
element $M_{12}^M$ by
 \beq
 \Delta m_M\approx 2|M_{12}^M|=\frac{1}{m_M}\left|\la
  \overline{M^0}|{\cal H}_{\rm eff}(|\Delta F|=2)|M^0\ra\right|,
 \eeq
where $|\Delta F|=2$ represents $|\Delta B|=2$ for the $\Bmixing$
mixing and $|\Delta C|=2$ for the $\Dmixing$ mixing. For the D meson
system, the above relation is valid under the assumption of CP
conservation. The effective operators which contribute to $\Delta
F=2$ are
 \beq
 Q_1&=&\bar q'\gamma_\mu(1-\gamma_5)q\bar q'\gamma^\mu(1-\gamma_5)q,  \non\\
 Q_2&=&\bar q'(1-\gamma_5)q\bar q'(1-\gamma_5)q.
 \eeq
We only keep the operators at the tree level and more operators
would emerge if QCD corrections are taken into account.

It is noted that the transferred momentum square for t- and
s-channels are approximately equal, i.e. $P^2\approx m_M^2$ for
heavy meson system.

Now we are able to give the expressions for the mass difference
$\Delta m_M$. The unparticle physics contribution $\Delta m_M^{\Un}$
is given as
 \beq
 \Delta m_M^\Un=\frac{5}{3}\frac{f_M^2\hat B_M}{m_M}\frac{A_{d_\Un}}{2|{\rm
 sin}d_\Un\pi|}\left(\frac{m_M}{\Lambda_\Un} \right)^{2d_\Un}|c_S|^2,
 \eeq
for the scalar unparticle and
 \beq
 \Delta m_M^\Un=\frac{f_M^2\hat B_M}{m_M}\frac{A_{d_\Un}}{2|{\rm
 sin}d_\Un\pi|}\left(\frac{m_M}{\Lambda_\Un} \right)^{2d_\Un-2}|c_V|^2.
 \eeq
for the vector unparticle. Note that in the above expression only
the absolute value of the function sin$d_\Un\pi$ exists. Our results
are the same as in \cite{CG} and slightly different from \cite{LZ}
by a constant factor. In the above derivations, we have used the
relations listed below \cite{BS}
 \beq
 \la \overline{M^0}|\bar q'\gamma_\mu(1-\gamma_5)q\bar q'\gamma^\mu(1-\gamma_5)q
  |M^0\ra=\frac{8}{3}f_M^2m_M^2\hat B_M, \non \\
 \la \overline{M^0}|\bar q'(1-\gamma_5)q\bar q'(1-\gamma_5)q
  |M^0\ra=-\frac{5}{3}f_M^2m_M^2\hat B_M.
 \eeq
where $f_M$ denotes the decay constant and $\hat B_M$ is a
numerical factor which is related to the non-perturbative QCD and
takes different values in various models, but as known, is of
order of unity.

Some comments are in order:

(1) The mass difference is proportional to a meson mass dependent
factor $m_M^{2d_\Un}$ or $m_M^{2d_\Un-2}$ which comes from the
unparticle propagator $\frac{1}{(P^2)^{2-d_\Un}}$. This is a
peculiar effect caused unparticle physics. The propagator for a
heavy particle exchange from other new physics does not depend on
the low energy scale $m_M$ in general.

(2) The above analysis is applicable to $\Bmixing$,
$B_s^0-\overline{B_s^0}$ and $\Dmixing$ mixing. For the K-system,
there are large uncertainties due to long-distance effects and the
approximations which exist in the theoretical calculations. Thus we
will not use the data on $K^0-\overline{K^0}$ mixing to constrain
the unparticle physics parameters.

(3) In this work, following the method commonly adopted  in
literature to study new physics effects, we assume that the new
physics beyond the SM which contributes to the mixing is the
unparticle sector. One can write
 \beq \Delta m_{M}^{NP}=\Delta
 m_{M}^{exp}-\Delta m_{M}^{SM},
 \eeq
where $\Delta m_{M}^{NP}$ corresponds to the contribution of new
physics, i.e. the unparticle in this study. The SM prediction on
$\Delta m_{B}$ has already been precise to two-loop order, and the
data are much more accurate than before thanks to the progress in
experimental measurements at Babar and Belle. Therefore by the
deviation between the SM prediction and measured value, we can set a
constraint on the parameters for the unparticle scenario.

Considering an extreme case, let us loosen the above restriction,
namely, we postulate that  the mixing $\Bmixing$ is fully due to the
unparticle contribution and see what constraints we would obtain on
the parameters. Later we will show that such constraints are looser
than that from that obtained from $\Dmixing$ mixing. Therefore, one
may not need to take the constraint on the unparticle parameters
from the data of $\Bmixing$ mixing at all.

(4) Because $\frac{\Delta m_{B_s}}{\Delta m_{B_d}}=34\gg 1$,
$B_s^0-\overline{B_s^0}$ mixing provides a looser constraint
compared to the $\Bmixing$ case.

The unknown parameters about the unparticles are: $\Lambda_\Un$,
$d_\Un$ and $c_S(c_V)$. In the numerical results, we fix the value
of $\Lambda_\Un$ by $\Lambda_\Un=1$ TeV. Other input parameters are:
$f_B\sqrt{\hat B}_B=0.2$ GeV \cite{Buras}, $f_D\sqrt{\hat B}_D=0.2$
GeV \cite{BS}, $\Delta m_{B_d}=0.507 ~ps^{-1}$ \cite{PDG}. The
recent experiment carried out by the Belle collaborations sets
$x_D=\frac{\Delta m_D}{\Gamma_D}=(0.80\pm 0.29(stat.)\pm
0.17(syst.))\%$ for the $\Dmixing$ \cite{Belle}. We use
$x_D<10^{-2}$ as the upper bound.

At first, we consider the case with $d_\Un=3/2$ and constrain $c_S$
and $c_V$ from $\Bmixing$ and $\Dmixing$ mixing.  Table \ref{t1}
lists the upper bounds for the coupling parameters $c_S$ and $c_V$.
The bounds obtained from $\Dmixing$ are more stringent than that
from $\Bmixing$ especially for the vector coupling $c_V$. This
confirms our expectation in the Introduction. The bounds obtained
from $\Dmixing$ mixing are: $|c_S|<2.1\times 10^{-2}$ and
$|c_V|<5.0\times 10^{-4}$.

\begin{table}[!h]
\caption{The upper bounds of $|c_S|$ and $|c_V|$ with
$\Lambda_\Un=1$ TeV and $d_\Un=3/2$. }\label{t1}
\begin{ruledtabular}
\begin{tabular}{ccc}
           & From B-system        & From D-system    \\ \hline
  $|c_S|$  & $3.4\times 10^{-2}$  & $2.1\times 10^{-2}$   \\ \hline
  $|c_V|$  & $2.3\times 10^{-3}$  & $5.0\times 10^{-4}$
\end{tabular}
\end{ruledtabular}
\end{table}

\begin{figure}[!htb]
\begin{center}
\begin{tabular}{cc}
\includegraphics[width=10cm]{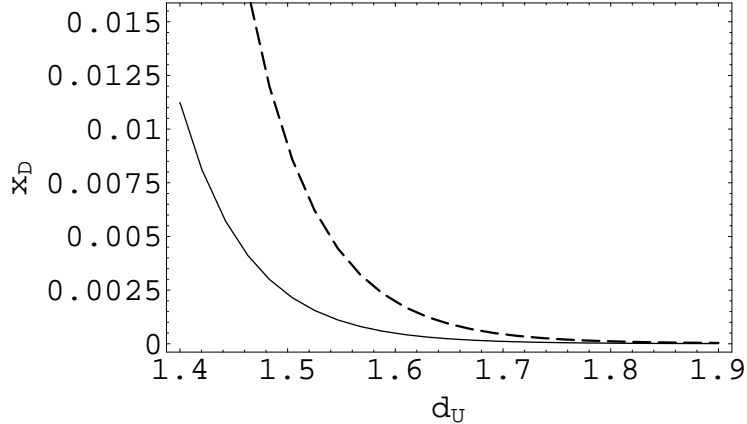}
\end{tabular}
\end{center}
\label{figure} \caption{ The $\Dmixing$ mixing parameter $x_D$
versus unparticle scale dimension ($1<d_\Un<2$). The solid line is
given for $|c_S|=1\times 10^{-2}$ and the dashed line for
$|c_S|=2\times 10^{-2}$.} \label{fig2}
\end{figure}

\begin{figure}[!htb]
\begin{center}
\begin{tabular}{cc}
\includegraphics[width=10cm]{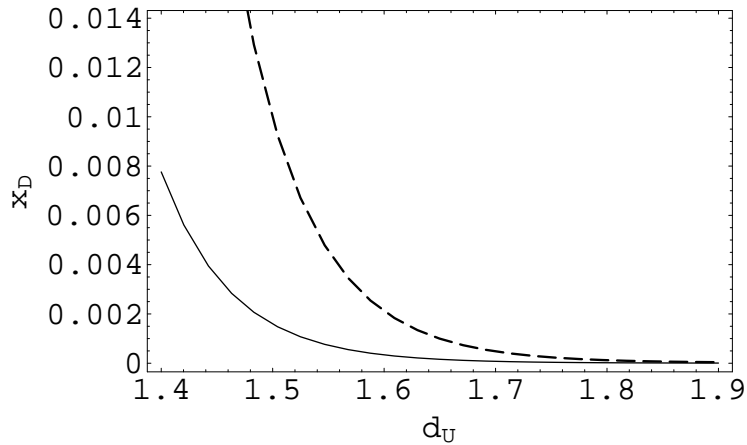}
\end{tabular}
\end{center}
\label{figure} \caption{ The $\Dmixing$ mixing parameter $x_D$
versus unparticle scale dimension ($1<d_\Un<2$). The solid line is
given for $|c_V|=2\times 10^{-5}$ and the dashed line for
$c_V=5\times 10^{-5}$.} \label{fig3}
\end{figure}

Then we consider the case with fixed $c_S$, $c_V$ and study the
dependence of the $\Dmixing$ mixing parameter $x_D$ on the scale
dimension $d_\Un$. Figs. \ref{fig2} and \ref{fig3} plot the
depedence within the parameter range $1<d_\Un<2$. We find that $x_D$
is very sensitive to $d_\Un$ and decreases rapidly to zero as
$d_\Un$ increases.

\begin{figure}[!htb]
\begin{center}
\begin{tabular}{cc}
\includegraphics[width=10cm]{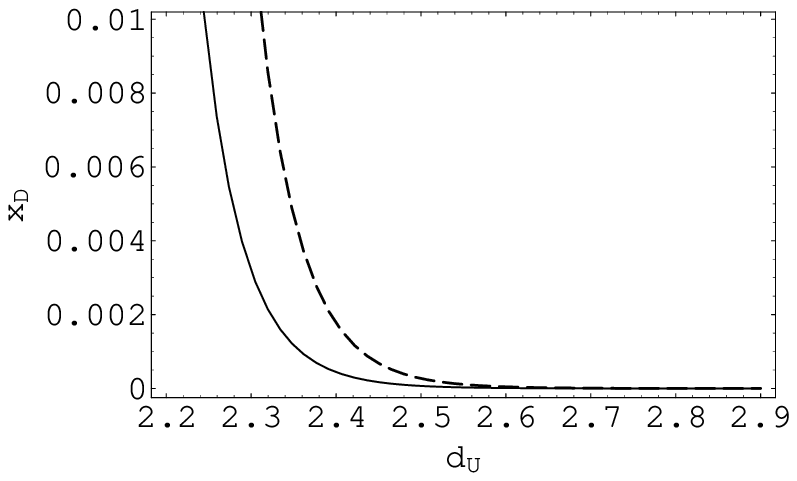}
\end{tabular}
\end{center}
\label{figure} \caption{ The $\Dmixing$ mixing parameter $x_D$
versus unparticle scaling dimension ($2<d_\Un<3$). The solid line is
given for $|c_S|=10$ and the dashed line for $|c_S|=20$.}
\label{fig4}
\end{figure}

\begin{figure}[!htb]
\begin{center}
\begin{tabular}{cc}
\includegraphics[width=10cm]{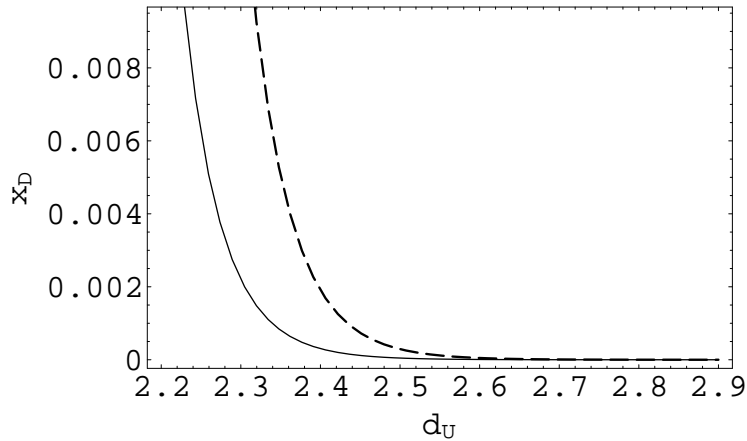}
\end{tabular}
\end{center}
\label{figure} \caption{ The $\Dmixing$ mixing parameter $x_D$
versus unparticle scaling dimension $2<d_\Un<3$. The solid line is
given for $|c_V|=2\times 10^{-2}$ and the dashed line for
$|c_V|=5\times 10^{-2}$.} \label{fig5}
\end{figure}

Moreover, we also investigate the case with extending the scale
dimension to the region $2<d_\Un<3$ and depict the dependence of
$x_D$ on $d_\Un $ in Figs. \ref{fig4} and \ref{fig5}. There is no
principal difference compared to the $1<d_\Un<2$ case except a
considerable change for the coupling parameters $c_S$ and $c_V$
which are required to fit the data.

\section{Conclusions}

We have investigated the new physics effects from scale invariant
unparticle sectors on the mixing of $\Bmixing$ and $\Dmixing$. The
exchange of unparticle induces the FCNC transitions at tree level
and provides new contribution to the mass difference of the meson
mass eigenstates. In principle, FCNC transitions may be caused by
other new physics effects which contain heavy massive particles and
break the scale invariance. We observe a peculiar effect caused by
the exchange of unparticle: the mixing parameter depends
non-trivially on the neutral meson mass. This dependence might not
occur for the heavy particle exchange from other new physics. We use
the data on $\Bmixing$ and $\Dmixing$ mixing to constrain the
parameters in unparticle scenario. We find that the $\Dmixing$
mixing provides the most stringent constraint on the coupling of the
scalar and vector unparticles to the SM quarks. The upper bounds we
obtained from $\Dmixing$ mixing are: $|c_S|<2.1\times 10^{-2}$ and
$|c_V|<5.0\times 10^{-4}$ if we set the energy scale $\Lambda_\Un=1$
TeV and scale dimension $d_\Un=3/2$. The dependence of scale
dimension $d_\Un$ shows that the mixing parameter is sensitive to
the scale dimension and decreases rapidly by almost two orders of
magnitude. The obtained parameters may have important effects on CP
violation in B and D decays.

\section*{Acknowledgments}

Z. Wei would like to thank Guo-Huai Zhu and Chuan-Hung Chen for
valuable discussions. This work was supported in part by NNSFC
under contract No. 10475042.

\end{document}